\documentclass[runningheads]{llncs}
\usepackage{graphicx}
\usepackage{booktabs}
\usepackage{hyperref}
\usepackage{listings}
\usepackage{url}
\usepackage{makeidx}
\usepackage{graphicx}
\usepackage{paralist}
\usepackage{multirow}
\usepackage[usenames, dvipsnames]{color}
\usepackage{wrapfig}
\usepackage{multicol}
\usepackage{wrapfig}
\graphicspath{ {images/} }
\usepackage[colorinlistoftodos]{todonotes}

\begin{document}
\title{Verification and Validation of Semantic Annotations}
\author{Oleksandra Panasiuk, Omar Holzknecht, Umutcan \c{S}im\c{s}ek, Elias K\"arle and Dieter Fensel }
\authorrunning{O. Panasiuk, O. Holzknecht, U. \c{S}im\c{s}ek, E. K\"arle and D. Fensel }
\institute{University of Innsbruck, Technikerstrasse 21a, Innsbruck 6020, Austria,\\ 
\email{firstname.lastname@sti2.at}}
\maketitle              % typeset the header of the contribution
\begin{abstract}
In this paper, we propose a framework to perform verification and validation of semantically annotated data. The annotations, extracted from websites, are verified against the schema.org vocabulary and Domain Specifications to ensure the syntactic correctness and completeness of the annotations. The Domain Specifications allow checking the compliance of annotations against corresponding domain-specific constraints. The validation mechanism will detect errors and inconsistencies between the content of the analyzed schema.org annotations and the content of the web pages where the annotations were found.

\keywords{verification \and  validation \and semantic annotation \and schema.org.}
\end{abstract}

\section{Introduction}
\label{sec:Introduction}
The introduction of the Semantic Web \cite{berners2001semantic} changed the way content, data and services are published and consumed online fundamentally.
For the first time, data in websites becomes not only machine-readable, but also machine understand- and interpretable. The semantic description of resources is driving the development of a new generation of applications, like intelligent personal assistants and chatbots, and the development of knowledge graphs and artificial intelligence applications.
The use of semantic annotations was accelerated by the introduction of schema.org \cite{guha2011introducing}. Schema.org was launched by the search engines Bing, Google, Yahoo! and Yandex in 2011. It has since become a de-facto standard for annotating data on the web \cite{mika2015schema}. The schema.org vocabulary,  serialized with Microdata, RDFa, or JSON-LD, is used to mark up website content. Schema.org is the most widespread vocabulary on the web, and is used on more than a quarter of web pages \cite{guha2016schema,meusel2014webdatacommons}.

Even though studies have shown that the amount of semantically annotated websites are growing rapidly, there are still shortcomings when it comes to the quality of annotations \cite{kaerle2016there,muhleisen2012web}. Also the analyses in \cite{hollenstein2016inconsistency,akbar2017complete} underline the inconsistencies and syntactic and semantic errors in semantic annotations. The lack of completeness and correctness of the semantic annotations makes content unreachable for automated agents, causes incorrect appearances in knowledge graphs and search results, or makes crawling and reasoning less effective for building applications on top of semantic annotations. These errors may be caused by missing guidelines, insufficient expertise and technical or human errors. Data quality is a critical aspect for efficient knowledge representation and processing. Therefore, it is important to define methods and techniques for semantic data verification and validation, and to develop tools which will make this process efficient, tangible and understandable, also for non-technical users.

In this paper, we extend our previous work \cite{csimcsek2017domain}, where we introduced a Domain Specification, and present an approach for verification and validation of semantic annotations. A Domain Specification (DS) is a design pattern for semantic annotations; an extended subset of types, properties, and ranges from schema.org. 
% a developed tool
The semantify.it Evaluator\footnote{\url{https://semantify.it/evaluator}}is a developed tool that allows the verification and validation of schema.org annotations which are collected from web pages. Those annotations can be verified against the schema.org vocabulary and Domain Specifications. The verification against Domain Specifications allows for the checking of the compliance of annotations against corresponding domain-specific constraints. The validation approach extends the functionality of the tool by detecting the consistency errors between semantic annotations and annotated content.

The remainder of this paper is structured as follows: Section~\ref{sec:Verification} describes the verification approach of semantic annotations. Section~\ref{sec:Validation} describes the validation approach. Section~\ref{sec:Conclusion } concludes our work and describes future work.

\section{Verification}
\label{sec:Verification}
In this section we discuss the verification process of semantic annotations according to schema.org and Domain Specifications. The section is structured as follows: Section~\ref{subsec:VerificationDefinition} gives the definition of the semantic annotation verification, Section~\ref{subsec:VerificationRelatedWork} describes related work, section~\ref{subsec:VerificationApproach} discusses our approach, and Section~\ref{subsec:VerificationEvaluation} describes the evaluation method. 

\subsection{Definition}
\label{subsec:VerificationDefinition}

The verification process of semantic annotations consists of two parts, namely, (I) checking the conformance with the schema.org vocabulary, and (II) checking the compliance with an appropriate Domain Specification. While the first verification step ensures that the annotation uses proper vocabulary terms defined in schema.org and its extensions, the second step ensures that the annotation is in compliance with the domain-specific constraints defined in a corresponding DS.

\subsection{Related Work}
\label{subsec:VerificationRelatedWork}

In this section, we refer to the existing approaches and tools to verify structured data. There are tools for verifying schema.org annotations, such as the Google Structured Data Testing tool\footnote{\url{https://search.google.com/structured-data/testing-tool/}}, the Google Email Markup Tester\footnote{\url{https://www.google.com/webmasters/markup-tester/}}, the Yandex Structured Data Validator\footnote{\url{https://webmaster.yandex.com/tools/microtest/}}, and 
the Bing Markup Validator \footnote{\url{https://www.bing.com/toolbox/markup-validator}}. They verify annotations of web pages that use Microdata, Microformats, RDFa, or JSON-LD as markup formats against schema.org. 
But these tools do not provide the check of completeness and correctness. For example, they can allow one to have empty range values, redundancy of information, or semantic consistency issues (e.g. the end day of the event is earlier than the start day).
In \cite{furber2010using} SPARQL and SPIN are used for constraint formulation and data quality check. The use of SPARQL and SPIN query template sets allows the identification of syntax errors, missing values, unique value violations, out of range values, and functional dependency violations.
The Shape Expression (ShEx) definition language \cite{prud2014shape} allows RDF verification\footnote{Authors use term "validation" in their paper due to content definition.} through the declaration of constraints. In \cite{boneva2014validating} authors define a schema formalism for describing the topology of an RDF graph that uses regular bag expressions (RBEs) to define constraints. 
In \cite{boneva2017semantics} the authors described the semantics of Shapes Schemas for RDF, and presented two algorithms for the verification of an RDF graph against a Shapes Schema. 
The Shapes Constraint Language\footnote{\url{https://www.w3.org/TR/shacl-ucr/}} (SHACL) is a language for formulating structural constraints on RDF graphs. SHACL allows us to define constraints targeting specific nodes in a data graph based on their type, identifier, or a SPARQL query.
The existing approaches can be adapted for our needs but not fully, as they are developed for RDF graph verification and not for schema.org annotations in particular.  

\subsection{Our approach}
\label{subsec:VerificationApproach}

To enable the verification of semantic annotations according to the schema.org vocabulary and to Domain Specifications, we developed a tool that executes a corresponding verification algorithm. This tool takes as inputs the schema.org annotation to verify and a DS that corresponds to the domain of the annotation. The outcome of this verification process is provided in a formalized, structured format, to enable the further machine processing of the verification result. 

The verification algorithm consists of two parts, the first checks the general compliance of the input annotation with the schema.org vocabulary, while the latter checks the domain-specific compliance of the input annotation with the given Domain Specification. The following objectives are given for the conformity verification of the input annotation according to the schema.org vocabulary:
\begin{enumerate}
  \item The correct usage of serialization formats allowed by schema.org, hence RDFa, Microdata, or JSON-LD.
  \item The correct usage of vocabulary terms from schema.org in the annotations, including types, properties, enumerations, and literals (data types).
  \item The correct usage of vocabulary relationships from schema.org in the annotations, hence, the compliance with domain and range definitions for properties.
\end{enumerate}

The domain-specific verification of the input annotation is enabled through the use of Domain Specifications\footnote{List of available Domain Specifications: \url{https://semantify.it/domainSpecifications/public}}, e.g. DS for annotation of tourism domain and GeoData  \cite{panasiuk2017defining,panasiuk18geodata}. Domain Specifications have a standardized data model. This data model consists of the possible specification nodes with corresponding attributes that can be used to create a DS document (e.g. specification nodes for types, properties, ranges, etc.). A DS document is constructed by the recursive selection of these grammar nodes, which, as a result, form a specific syntax (structure) that has to be satisfied by the verified annotations \cite{mastersthesisOmar2018}. Keywords in these specification nodes allow the definition of additional constraints (e.g. "multipleValuesAllowed" or "isOptional" for property nodes). In our approach, the verification algorithm has to ensure that the input annotation is in compliance with the domain-specific constraints defined by the input DS. In order to achieve this, the verification tool has to be able to understand the DS data model, the possible constraint definitions, and to check if verified annotations are in compliance with them.

\subsection{Evaluation}
\label{subsec:VerificationEvaluation}

\begin{figure}
\centering
	\includegraphics[width=0.9\linewidth]{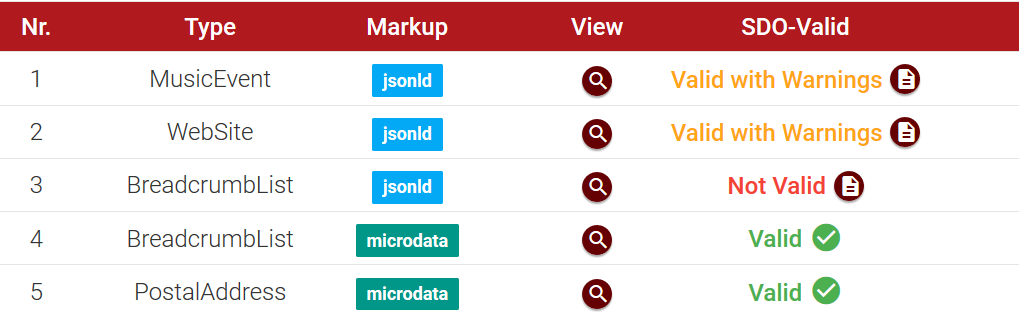}
	\caption{Schema.org Verification}
	\label{fig:SDOValidation}
\end{figure}

\begin{figure}
\centering
	\includegraphics[width=0.9\textwidth]{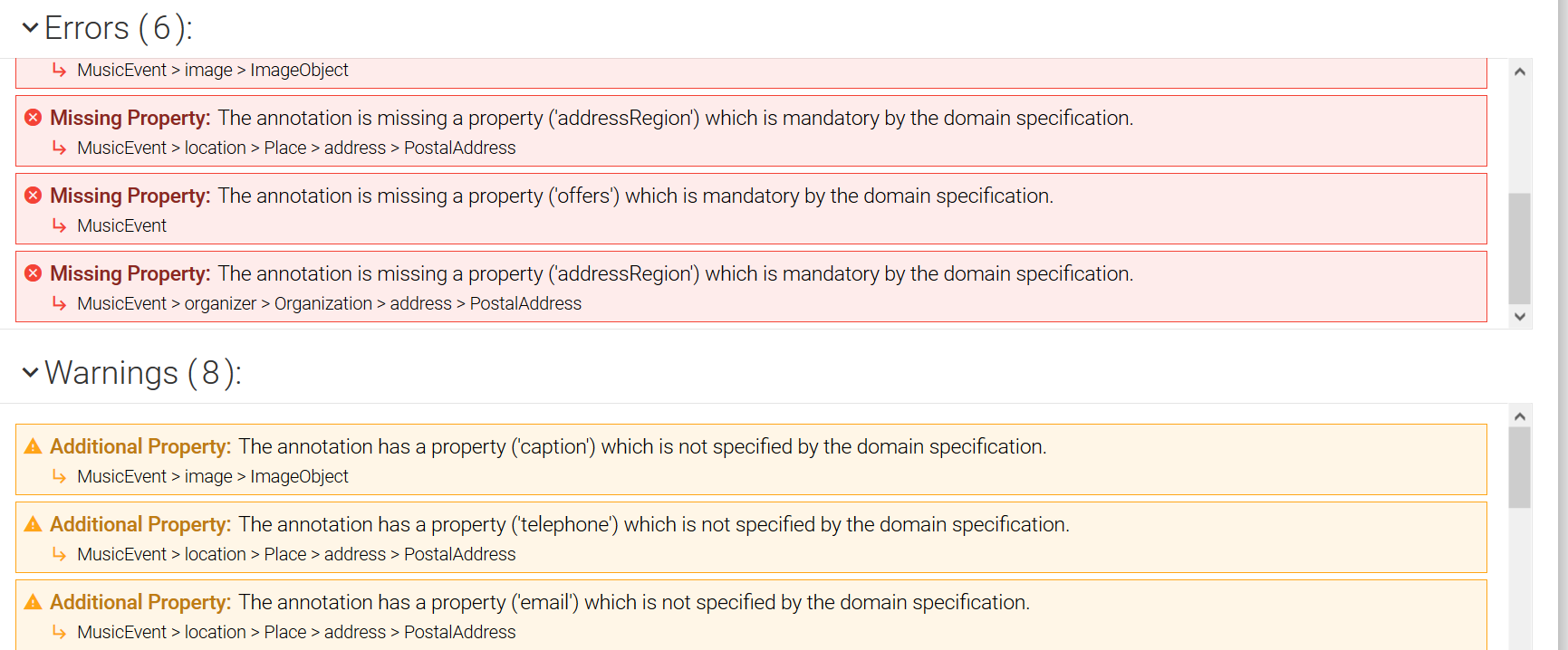}
\caption{Domain Specification Verification. Verification Report}
\label{fig:DSValidation1}
\end{figure}

We implement our approach in the semantify.it Evaluator\footnote{\url{https://semantify.it/evaluator}}.
The tool provides a verification report with detailed information about detected errors according to the schema.org vocabulary (see Fig.\ref{fig:SDOValidation}) and  Domain Specifications (see  Fig.\ref{fig:DSValidation1}).

%Figure 1

%Figure 2

\begin{figure}
\centering
	\includegraphics[width=0.9\textwidth]{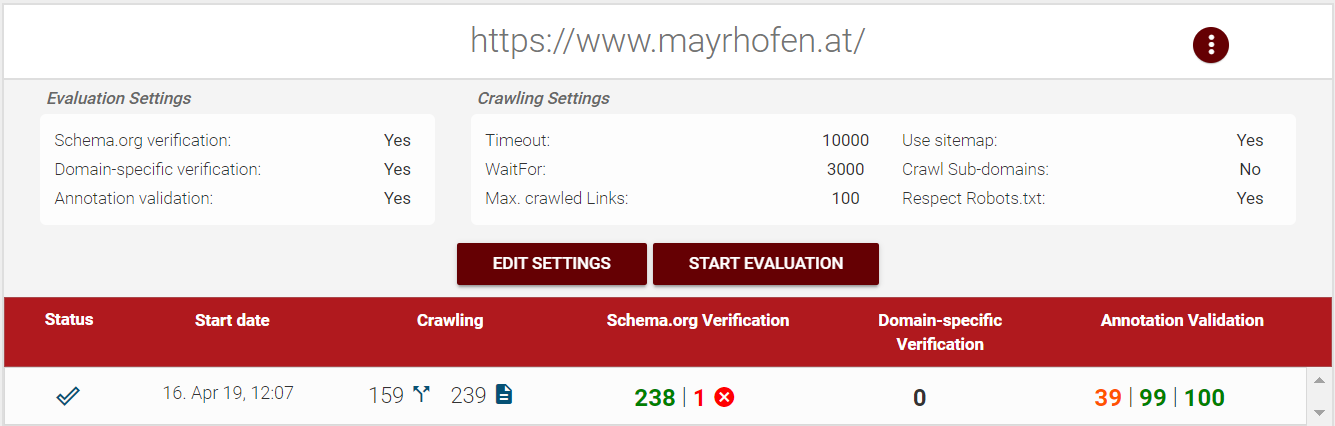}
\caption{semantify.it Evaluator. Verification and Validation Report}
\label{fig:ReportEvaluator}
\end{figure}

Besides the verification result itself, the report includes details about the detected errors, e.g. error codes (ID of the error type), error titles, error severity levels, error paths (where within the annotation the error occurred), and textual descriptions of the errors.
The implementation itself can be evaluated through unit tests in terms of a correct functionality (correctness) and the implementation of all possible constraint possibilities of the Domain Specification vocabulary (completeness).
This can be achieved by comparing the structured representation of the result, namely the JSON file produced by the verification algorithm, which is used to generate a human-readable verification report for the user (see Fig.\ref{fig:ReportEvaluator}), with the expected verification report outcome specified in the test cases for predefined annotation-Domain Specification pairs.

%Figure 3

A formal proof of the correctness and completeness of our implemented algorithm is rather straight forward given the simplicity of our current knowledge representation formalism. In our ongoing work\footnote{The paper is under double blind review and can't be revealed}, we develop a richer constraint language which will require more detailed analysis of these issues.

\section{Validation}
\label{sec:Validation}
Search engines may penalize the publisher of structured data if their annotations include content that is invisible to users, and/or markup irrelevant or misleading content. These penalties may have negative effects on a website (e.g. bad position of the website in search results) or even lead to a non-integration of the structured data (e.g. no generation of rich snippets). For example, annotations of the Destination Management Organizations (DMOs) usually include a list of offers. These offers must comply with offers which are described on the website, and all URLs contained in the annotations must match with the URLs in the content. Such issues can be detected through the validation of semantic annotations.  

In this section, we discuss the validation process of semantic annotations and the proposed approach.
The section is structured as follows: Section~\ref{subsec:ValidationDefinition} gives the definition of the semantic annotation validation, Section~\ref{subsec:ValidationRelatedWork} describes some related work, Section~\ref{subsec:ValidationApproach} discusses our approach, and Section~\ref{subsec:ValidationEvaluation} describes the evaluation method. 

\subsection{Definition}
\label{subsec:ValidationDefinition}
The validation of semantic annotations is the process of checking whether the content of a semantic annotation corresponds to the content of the web page that it represents, and if it is consistent with it. Semantic annotations should include the actual information of the web page, correct links, images and literal values without overlapping or redundancy. 

\subsection{Related Work}
\label{subsec:ValidationRelatedWork}
The incorrect representation of the structured data can make data unreachable for automated engines, cause an incorrect appearance in the search results, or make crawling and reasoning less effective for building applications on top of semantic data.  
The errors may be caused by not following recommended guidelines, e.g. structured data guidelines\footnote{\url{https://developers.google.com/search/docs/guides/sd-policies}}, insufficient expertise, technical or human errors (some of the issues can be detected by Google search console\footnote{\url{https://search.google.com/search-console/about}}), and/or annotations not being in accordance with the content of web pages, so-called "spammy structured markup"\footnote{\url{https://support.google.com/webmasters/answer/9044175?hl=en&visit_id=636862521420978682-2839371720&rd=1\#spammy-structured-markup}}.
There is no direct literature related to the methods of detecting inconsistency between semantic annotations and content of web pages, but the problem of the content conformity restriction is also mentioned in \cite{karle2018heuristics}.

\subsection{Our approach}
\label{subsec:ValidationApproach}
Since semantic annotations are created and published by different data providers or agencies in varying quantity and quality and using different assumptions, the validity of data should be prioritized to increase the quality of structured data. To solve the problem of detecting errors caused by inconsistencies between analyzed schema.org annotations and the content of the web pages where the annotations were found, we propose a validation framework. The framework consists of the following objectives:
\begin{enumerate}
    \item Detect the main inconsistencies between the content of schema.org annotations and the content of their corresponding web pages. 
    
    \item Develop an algorithm for the consistency check between a web page and corresponding semantic annotations. 
    The information from web pages can be extracted from the source of a  web page by tracking the appropriate HTML tags, keywords, lists, images, URLs, paragraph tags and the associated full text. Some natural language processing and machine learning techniques can be applied to extract important information from the textual description, e.g price, email, telephone number and so on. There exist some approaches, such as named entity recognition \cite{mohit2014named} to locate and categorize important nouns and proper nouns in a text, web information extraction systems \cite{chang2006survey}, text mining techniques \cite{allahyari2017brief}.
    
    \item Define metrics to evaluate the consistencies of the semantic annotations according to the annotated content. 
    %There are different metrics which can be applied to the validation process. They were formulated in the literature review of quality methodologies \cite{batini2009methodologies,zaveri2012quality}, models and metrics for big data and linked open data quality assessment \cite{cai2015challenges,knight2005developing,pipino2002data}, ontology foundation and evaluation  \cite{wand1996anchoring,hlomani2014approaches}, ontology design patterns \cite{hammar2017contentODP}, knowledge graphs \cite{farber2018linked}, and semantic description of web services \cite{zhu2017quality}. 
    In this step, we analyze existing data quality metrics that can be applied on the structured data and define metrics that can be useful to evaluate the consistency between a web page content and semantic annotation. We measure the consistency for different types of values, such as URL, string, boolean, enumeration, rating value, date and time formats. 
    
    \item Provide a validation tool to present the overall score for a web page and detailed insights about the evaluated consistency scores on a per value level. 
\end{enumerate}

\subsection{Evaluation}
\label{subsec:ValidationEvaluation}
\begin{figure}
\centering
	\includegraphics[width=0.48\linewidth]{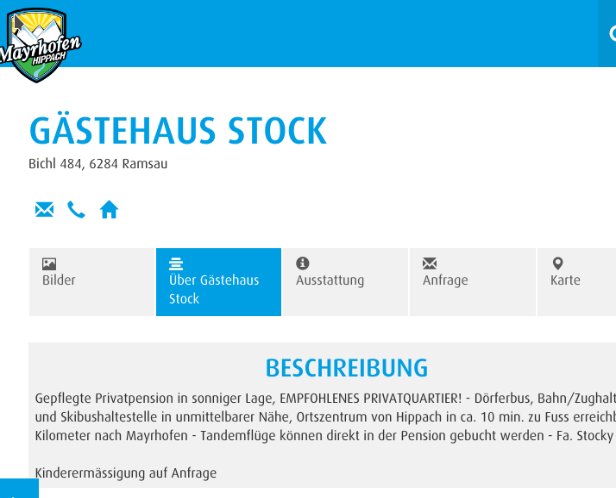}
	\includegraphics[width=0.47\linewidth]{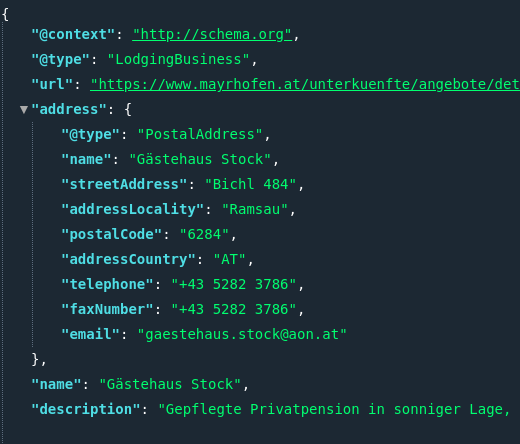}
	\caption{Web page content and annotation content}
	\label{fig:WP_Ann}
\end{figure}

To ensure the validity of the report results, we will organize a user study of semantic annotations and annotated web pages to prove the performance of our framework. The questionnaire will be structured in a way to get quantitative and qualitative feedback about the consistencies between a web page and annotation content (see Fig. \ref{fig:WP_Ann}) according to the results provided by the framework (see Fig.\ref{fig:ReportEvaluator}). As our use case, we will use annotated data and websites of Destination Management Organizations, such as Best of Zillertal F\"ugen\footnote{\url{https://www.best-of-zillertal.at}}, Mayrhofen\footnote{\url{https://www.mayrhofen.at}}, Seefeld\footnote{\url{https://www.seefeld.com/ }}, and Zillertal Arena\footnote{\url{https://www.zillertalarena.com }}.

\section{Conclusion and Future Work }
\label{sec:Conclusion }
Semantic annotations will be used for improved search results by search engines or as building blocks of knowledge graphs. Therefore, the quality issues in terms of structure and consistency can have an impact on where the annotations are utilized and lead, for instance, to false representation in the search results or to low-quality knowledge graphs.
In this paper, we described our ongoing work for an approach to verify and validate semantic annotations and the tool that is evolving as the implementation of this approach. 
%The current state of the approach can verify annotations in terms of structure and syntax based on Domain Specifications. In our ongoing we are extending our approach and the tool with validation, which means to check whether the annotation is consistent with the content that it describes. This is especially important for search engine results, since inconsistency between content and annotation may be perceived as spam and harm the search result rankings.

For the future work, we will define Domain Specifications with SHACL in order to comply with the recent W3C Recommendation for RDF validation. We will develop an abstract syntax and formal semantics for Domain Specifications and map it to SHACL notions, for instance by aligning the concept of Domain Specifications with SHACL node shapes.

\bibliographystyle{splncs04}
\bibliography{bib}

\end{document}